\documentclass{article}
\usepackage[T1]{fontenc}
\usepackage{geometry}
\usepackage{setspace}
\makeatletter

\providecommand{\LyX}{L\kern-.1667em\lower.25em\hbox{Y}\kern-.125emX\@}
\let\SF@@footnote\footnote
\def\footnote{\ifx\protect\@typeset@protect
    \expandafter\SF@@footnote
  \else
    \expandafter\SF@gobble@opt
  \fi
}
\expandafter\def\csname SF@gobble@opt \endcsname{\@ifnextchar[
  \SF@gobble@twobracket
  \@gobble
}
\edef\SF@gobble@opt{\noexpand\protect
  \expandafter\noexpand\csname SF@gobble@opt \endcsname}
\def\SF@gobble@twobracket[#1]#2{}

 \newcommand{\lyxaddress}[1]{
   \par {\raggedright #1 
   \vspace{1.4em}
   \noindent\par}
 }

\makeatother

\begin{document}

\title{{\large The condition~ 0 < Z < 1~ and an intrinsic mass scale in Quantum
Field Theory}\large }

\author{Satish D. Joglekar{*}}

\maketitle

\lyxaddress{{\large Department of Physics }\large }

\lyxaddress{{\large Indian Institute of Technology, Kanpur }\large }

\lyxaddress{{\large 208016, U.P., India}\large }

\begin{abstract}
{\large In this work, we suggest a view-point that leads to an intrinsic mass
scale in Quantum Field Theories.This view-point is fairly independent of dynamical
details of a QFT and does not rely on any particular framework to go beyond
the standard Model.We use the setting of the nonlocal quantum field theories
{[}NLQFT{]} with a finite scale parameter~ \( \Lambda  \), which are unitary
for finite \( \Lambda  \). We propose that the condition 0 < Z~ < 1 {[}wherever
proven{]} can be rigorously implemented/imposed in such theories and that it
implies the existence of a mass scale \( \Lambda  \)~ that can be determined
from this condition. We derive the nonlocal analogue of the above relation {[}which
is a} \emph{\large finite} {\large relation in NLQFT{]} and demonstrate that
it can be arrived at only from general principles. We further propose that the
nonlocal formulation should be looked as an effective field theory that incorporates
the effect of dynamics beyond an energy scale and which breaks at the intrinsic
scale \( \Lambda  \) so obtained. Beyond this scale it should be replaced by
another {[}perhaps, a more fundamental{]} theory. We provide a heuristic justification
for this view-point.}{\large \par}
\end{abstract}
{\large {*}email: sdj@iitk.ac.in}{\large \par}

\section{\textmd{\large INTRODUCTION}\large }

\paragraph{\textmd{\large ~~~~~~~~Forthcoming data from accelerators will enable
one to access energies in the TeV region and will ,in addition, to testing the
Standard Model ,enable to discover Physics at the higher energy scales of TeV
and beyond.There are a number of models which predict new Physics beyond the
Standard Model. These are based on various schemes of Grand Unification/intermediate
scale unification as well as models based on Supersymmetry and superstrings.These
works and recent works involving higher dimensions predict scales of energies
at which new Physics would emerge.These mass scales depend, naturally, on the
dynamical details of the models.In this work,we wish to suggest a view-point,
based solely on a given Quantum field theory used for describing physical processes
at present energies.We wish to suggest that an intrinsic scale exists in a Quantum
Field Theory and that it can lead to testable consequences in the near future.A
view point, similar to this, but based on the understanding of the renormalization
program,has also been suggested in Ref.12.This view-point is likely to be of
a general use without specific reference to a model used to predict physics
beyond the Standard Model.}\large }

\paragraph{\textmd{\large The presently successful theory of strong, electromagnetic and
weak interactions, viz. the standard model (SM) is a Local Quantum Field Theory
(LQFT) {[}1{]}.~ Local Quantum Field Theory calculations, when done perturbatively
are generally plagued with divergences and this certainly holds for the SM calculations
{[}2{]}.~ The initial successes of the first LQFT viz. Quantum electrodynamics
(QED) were based upon the renormalization program initiated by Feynman, Schwinger,
Tomonaga and developed to a general form by Dyson {[}3{]}.~ This program gives
an elaborate procedure for dealing with divergences in LQFT.~ When this procedure
is followed order by order in perturbation theory, it was demonstrated that
all the divergences in the theory could be absorbed in the definitions of renormalized
fields and parameters as related to the unrenormalized ones.~ These relations
were obtained by imposing by hand \char`\"{}renormalization prescriptions\char`\"{}
on the 1PI (proper) vertices which amounted to giving by hand (i.e., from experiment)
the physical masses and couplings (and other unphysical parameters) then the
renormalized S-matrix was indeed finite in terms of these.~ This procedure
was highly successful for QED and more so for the further development of Standard
Model {[}2{]}.~ It also yielded many results based on renormalization group
equations and Callan-Symanzik equation {[}4{]}.}\large }

{\large ~}{\large \par}

\paragraph{\textmd{\large ~~~~~~~~The renormalization procedure, despite several
initial misgivings, came to be regarded as an essential established part of
LQFT primarily due to the successes of renormalized LQFT in particle physics.~
However, as any text book discussion shows, the treatment of divergences in
perturbation theory is highly suspicious from the point of view of mathematical
rigor {[}See e.g. Ref.2{]}.}\large }

{\large ~}{\large \par}

\paragraph{\textmd{\large ~~~~~~~~Definition of the infinite Feynman integrals
involved requires a regularization.~ A regularization such as Pauli-Villars
violates unitarity for any finite cut off {[}5{]}, which is recovered only as
\protect\( \Lambda \protect \) -> \protect\( \infty \protect \).~ Further
in a calculation to any finite order of perturbation theory one makes mathematically
unjustified truncations.~ Thus, in a Pauli-Villars regularization, if }\large }

{\large ~}{\large \par}

{\large ~~~~~~~~~~~~~~~~Z = 1 + a g\( ^{2} \) ln \( \Lambda  \)\( ^{2} \)
+O(g\( ^{4} \))~~~~~~~~~~~~~~~~~~~~~~~~~~~~~~~~~~~~~~~~~~~~~~~~~~~~~~~~~~~~~~~~~
(1.1)}{\large \par}

{\large ~}{\large \par}

{\large Z\( ^{-1} \) is truncated to}{\large \par}

{\large ~}{\large \par}

{\large ~~~~~~~~~~~~~~~~Z\( ^{-1} \) = 1 - a g\( ^{2} \) ln
\( \Lambda  \)\( ^{2} \) +O(g\( ^{4} \))~~~~~~~~~~~~~~~~~~~~~~~~~~~~~~~~~~~~~~~~~~~~~~~~~~~~~~~~~~~~~~~~~
(1.2) }{\large \par}

{\large which is mathematically invalid even for finite (but large enough) \( \Lambda  \).~
Similar truncations are made in any regularization.~ Thus one does not have,
in the conventional renormalization of LQFT procedure, unitarity and mathematical
consistency for any finite (but large enough)     \( \Lambda  \).~ Further,
the relation(s)such as 0 < Z < 1 for the wavefunction renormalizations (wherever
applicable) obtained from LSZ formulation without recourse to perturbation procedure
{[}6{]} have to be ignored in these procedures as Z turns out to diverge in
perturbation theory {[}6{]}.~ Despite these mathematical shortcoming the renormalization
program has succeeded exceedingly well.}{\large \par}

\paragraph{\textmd{\large ~~~~~~~~Since early days, one has felt that it may be
possible to cure the procedure of these shortcomings; but it has not been possible.~
However, now nonlocal formulations of field theories (NLQFT) are possible {[}7,8{]}
in which the theories have a finite scale \protect\( \Lambda \protect \) and
are finite (with \protect\( \Lambda \protect \) finite), unitary and causal
for finite \protect\( \Lambda \protect \). We find it convenient to use such
a formulation as the background for our line of reasoning.~~ In such formulations,
gauge (and other symmetries) can also be preserved, in a generalized (nonlocal)
form {[}8{]}.~ They also admit results of renormalization group at finite \protect\( \Lambda \protect \).~
One can look upon these formulations in two possible ways: (i) as a new nonlocal
regularization scheme, an augmentation of the available regularization and renormalization
procedures or (ii) as theories in which \protect\( \Lambda \protect \) having
a fixed finite value serves as the underlying (possibly effective) theory itself.~
This latter view point has been proposed originally {[}in some form{]} in Ref.7,
stressed also in ref.8 and has also been extended and followed up in Ref. 9
and 12.~ In these theories, all calculations are (strictly) finite and (finite)
renormalization procedure is needed only for organization of calculations to
a given order.~ We wish to work in the context of such a theory with a finite
\protect\( \Lambda \protect \).~ We wish to demonstrate that in such formulations,
the mathematical consistency requirements and the relation 0 < Z < 1 can in
fact be implemented literally and nontrivial conclusions can be drawn from it,
which would not be possible in the usual formulation of the renormalization
procedure. This approach has, in fact, also been able to explain where and why
the usual renormalization procedure works {[}12{]}.}\large }

\paragraph{\textmd{\large ~~~~~~~~We outline the approach(es) we want to adopt.~
We suppose that the particle physics theories are in fact described by a nonlocal
action of the type proposed in Ref. {[}8{]} with a finite parameter \protect\( \Lambda \protect \)
present in it.~ Presence of such a parameter can be looked upon in two ways
{[}9{]}; and we shall discuss our results in the context of both.~ In the first
approach we may regard 1/\protect\( \Lambda \protect \) as the scale of nonlocality
arising possibly from a fundamental length scale already existing in nature.~
In this approach, the NLQFT is an exact theory valid to all energies.~ In the
second approach, which is probably more plausible, the nonlocal theory is looked
upon an effective field theory valid upto a certain energy scale (dependent
on \protect\( \Lambda \protect \)) and beyond this scale, the theory would
have to be replaced by another NLQFT of a more fundamental nature. We elaborate
on and justify this view further in Sec.3.}\large }

\paragraph{\textmd{\large ~~~~~~~~We summarize the plan of the paper.~ In section
II, we briefly introduce the nonlocal quantum field theories of Ref. {[}7,8{]}
which we use as the base for our arguments.~ We also formulate our interpretation
of these theories useful in this work. In Sec III, we deal with the relation
0 < Z < 1, and show that this can be implemented in perturbation theory meaningfully
and show it also implies the existence of a mass scale by itself. We also give
here a simple derivation of the analogous relation in NLQFT. We suggest an the
interpretation of NLQFT's and offer a justification for this view-point. We
make tentative calculations to show how we can arrive at mass scale \protect\( \Lambda \protect \).~
We conclude that we may~ end up with a scale whose effects may be testable
experimentally in near future. Moreover, with this mass scale, the usual perturbative
manipulations such as those in (1.1)-(1.2) may also turn out to be mathematically
rigorous{[}12{]}.It is possible that models based on non-commutative geometry
{[}which also involves a scale{]} could lead to a similar result, if they could
be developed to the same degree as the nonlocal field theories{[}13{]}.}\large }

{\large ~}{\large \par}

\section{\textmd{\large Nonlocal Quantum Field Theories (NLQFT) and their physical interpretation}\large }

{\large ~}{\large \par}

\paragraph{\textmd{\large ~~~~~~~~In this section, we shall introduce the way
of formulating nonlocal quantum field theories given in reference 8~ with the
help of \protect\( \phi ^{4}\protect \) theory and elaborate on physical interpretations
we wish to associate with it.}\large }

\subsection{\textmd{Nonlocal \protect\( \phi ^{4}\protect \) theory}}

\paragraph{\textmd{\large ~~~~~~~~Consider a local theory with the action written
as a sum of the free and interacting parts}\large }

\subparagraph{\textmd{\large S{[}\protect\( \phi \protect \){]} = \protect\( \frac{1}{2}\sum _{\phi }\int d^{4}x\protect \)~\{
\protect\( \phi F\phi \protect \)~+I{[}\protect\( \phi ]\}\protect \)~~~~~~~~~
~~~~~~~~~~~~~~~~~~~~~~~~~~~~~~~(2.1)}\large }

\subparagraph{\textmd{\large where \protect\( \varphi \protect \) represents the fields
(fermionic, bosonic) of the theory with the appropriate spacetime and internal
symmetry group indices.~ F is the 'kinetic' operator for the field \protect\( \varphi \protect \)
and I{[}\protect\( \varphi \protect \){]} is the interaction part of S{[}\protect\( \varphi \protect \){]}.~
For the \protect\( \phi ^{4}\protect \) theory,}\large }

\subparagraph{\textmd{\large ~~~~~~~~~~~~~~~~F = (-\protect\( \partial ^{2}\protect \)
- \protect\( \mu \protect \)\protect\( ^{2}\protect \)) ~~~~~~~~~~~~~~~~~~~~~~~~~~~~~~~~~~~~~~~~~~~~~~~~~~(2.2)}\large }

\subparagraph{\textmd{\large and}\large }

\subparagraph{\textmd{\large ~~~~~~~~I{[}\protect\( \phi ]\}\protect \)= \protect\( \int \protect \)\protect\( d^{4}x\frac{\lambda }{4}\protect \)\protect\( \phi ^{4}\protect \)~~~~~~~~
~~~~~~~~~~~~~~~~~~~~~~~~~~~~~~~~~~~~~~~~~~~~~~~(2.3)}\large }

\paragraph{\textmd{\large Nonlocalization of S{[}\protect\( \varphi \protect \){]} is
carried out using a 'smearing' operator defined in terms of the kinetic operator
F of the theory as}\large }

\paragraph{\textmd{\large ~~~~~~~~~~~\protect\( \varepsilon \protect \)~=
exp{[}\protect\( \frac{F}{2\Lambda ^{2}}]\protect \)~~~~ ~~~~~~~~~~~~~~~~~~~~~~~~~~~~~~~~~~~~~~~~~~~~~~~(2.4)}\large }

\paragraph{\textmd{\large \protect\( \Lambda \protect \) is the scale of nonlocality.~
With the help of the smearing operator, we define a smeared field~~ by}\large }

\paragraph{\textmd{\large ~~~~~~~\protect\( \widehat{\phi }\protect \)~= \protect\( \varepsilon \protect \)\protect\( ^{-1}\protect \)\protect\( \phi \protect \)
~~~~~~~~~~~~~~ ~~~~~~~~~~~~~~~~~~~~~~~~~~~~~~~~~~~~~~~(2.5)}\large }

\paragraph{\textmd{\large Next, for every field \protect\( \varphi \protect \) we introduce
an auxiliary 'shadow' field \protect\( \varphi \protect \)\protect\( ^{sh}\protect \)
of the same type as \protect\( \varphi \protect \) which couples to \protect\( \varphi \protect \)
through an auxiliary action S given by }\large }

\paragraph{\textmd{\large ~~~~S {[}\protect\( \phi ,\phi ^{sh}]=\protect \)~\protect\( \frac{1}{2}\sum _{\phi }\int d^{4}x\protect \)~\{
\protect\( \widehat{\phi }\protect \)\protect\( F\widehat{\phi }\protect \)~+\protect\( \phi ^{sh}\vartheta ^{-1}\phi ^{sh}+\protect \)I{[}\protect\( \phi +\phi ^{sh}]\}\protect \)~~~~~~~~~~~~~~(2.6)}\large }

\paragraph{\textmd{\large where \protect\( \vartheta \protect \)is the 'shadow' kinetic
operator defined as }\large }

\paragraph{\textmd{\large ~~~~~~~~~~~~~~~~\protect\( \vartheta \protect \)=(\protect\( \varepsilon ^{2}\protect \)
- 1)F\protect\( ^{-1}\protect \)~~~~~~~~~~~~~~~~~~~~~~~~~~~~~~~~~~~~~~~~~~~~~~~~~(2.7)}\large }

\paragraph{\textmd{\large The action for the nonlocalized theory \protect\( \widehat{S}[\phi ]\protect \)
is then defined by }\large }

\paragraph{\textmd{\large ~~~~~~~~~~~~~~\protect\( \widehat{S}[\phi ]\protect \)=S{[}\protect\( \phi ,\phi \protect \)\protect\( ^{sh}(\phi )]\protect \)~~
~~~~~~~~~~~~~~~~~~~~~~~~~~~~~~~~~~~~~~~~~~~~~~~(2.8)}\large }

\paragraph{\textmd{\large where \protect\( \varphi \protect \)\protect\( _{sh}\protect \){[}\protect\( \varphi \protect \){]}
is the solution of the classical shadow field equation }\large }

\paragraph{\textmd{\large ~~~~~~~~~~~~\protect\( \frac{\delta S}{\delta \phi ^{sh}}\protect \){[}\protect\( \phi ,\phi ^{sh}\protect \)
{]}=0~~~ ~~~~~~~~~~~~~~~~~~~~~~~~~~~~~~~~~~~~~~~~~~~~~~~(2.9)}\large }

\paragraph{\textmd{\large Quantization is carried out in the path integral formulation.~
The quantization rule is given by }\large }

\subparagraph{\textmd{\large <T{*}(O{[}\protect\( \phi ])>=\int \protect \)~D\protect\( \phi \mu [\phi ]O[\widehat{\phi ]}\protect \)
exp {[}i\protect\( \widehat{S}\protect \){[}\protect\( \phi ]]\protect \)~~~~~~~~~~~~
~~~~~~~~~~~~~~~~~~~~~~~~~~~~~~(2.10)}\large }

\paragraph{\textmd{\large These theories are known to be unitary even for finite \protect\( \Lambda \protect \),
and causal}\footnote{%
They have very weak violations of causality.
} \textmd{\large {[}8{]}.~ In such formulations, gauge and other symmetries
can be preserved in a generalized nonlocal form {[}8{]}.~ They also admit results
on renormalization group at finite \protect\( \Lambda \protect \) {[}8{]}.
These formulations have been proposed as regularizations of field theories for
\protect\( \Lambda \protect \)--> \protect\( \infty \protect \). More importantly,
there are alternate interpretations for theories with a finite \protect\( \Lambda \protect \);
these have already been expressed in Section I and will be the ones that interest
us in this work.}\large }

\section{\textmd{\large INTERPRETATION OF 0 < Z < 1 RELATION IN NLQFT's}\large }

\paragraph{\textmd{\large In this section, we wish to first of all point out that the
relation implied by the LSZ theory{[}11{]}, viz. 0 < Z < 1 {[}wherever such
a relation can be formulated{]}, can in fact be implemented in perturbation
theory contrary to what has become the prevalent belief; and further that it
can, in fact, have an interesting interpretation for the scale of validity of
a particular local Quantum Field Theory formulation as a physically viable theory.
{[}We note here that Z is the wavefunction renormalization on mass shell and
has no ambiguity about it.{]}}\large }

\paragraph{\textmd{\large The relation 0 < Z < 1 has a simple physical interpretation;
in that Z relates to the probability that the state created by the Quantum Field
operator \protect\( \varphi \protect \)(x) | 0> contains in it the single particle
states.In a free field theory, we expect none but the single particle states
in \protect\( \varphi \protect \)(x) |0> and hence, Z = 1.As a result of interaction,
multiparticle states become possible and the net probability~ of single particle
states in \protect\( \varphi \protect \)(x) | 0> diminishes.Of course, we never
expect this probability to be negative; hence 0 < Z.The relation }\large }

\subparagraph{\textmd{\large ~~~~~~~~~~~~~~~~~~ 1 = Z +\protect\( \int _{_{_{_{m^{2}_{1}}}}}^{\infty }\protect \)\protect\( \rho (\sigma \protect \)\protect\( ^{2})d\sigma \protect \)\protect\( ^{2}\protect \)}\large }

\subparagraph{\textmd{\large is the partition of the total probability into quantities related
to the single particle and the multiparticle scattering states.}\large }

\paragraph{\textmd{\large ~ ~~~~~~In QFT ,this physical interpretation does not
seem viable because \protect\( \int _{_{_{_{m^{2}_{1}}}}}^{\infty }\protect \)\protect\( \rho (\sigma \protect \)\protect\( ^{2})d\sigma \protect \)\protect\( ^{2}\protect \)
diverges in perturbation theory and consequently Z -->\protect\( \infty .\protect \)This
,of course, looks a suspect situation for QFT ;but when the renormalization
procedure was advanced and shown to work extremely well theoretically and to
lead to accurately verifiable experimental results {[}2{]}, physicists developed
faith in it and found it unavoidable to ignore this relation for Z. It has been
believed that we need not take this relation seriously in perturbatively, though
it might hold nonperturbatively.}\large }

\paragraph{\textmd{\large ~ ~~~~~~We wish to propose that the relation 0 < Z <
1 {[}wherever proven{]} can, in fact, be implemented in perturbation theory
in NLQFT's; have the physical meaning it is said to have in the earlier part
of this section and at the same time preserve the advantages of the standard
renormalization program{[}12{]} .{[}In other words, the proposal here does not
clash in any way with the standard renormalization program but supplements it.{]}
It is but natural that the implementation of {[}an extra {]} relation can lead
to a new physical output and we shall propose one in this section: An energy
scale !}\large }

\paragraph{\textmd{\large ~~ ~~~~~We find it fruitful to combine the last interpretation
of the NLQFT's given in the introduction and the relation~ 0 < Z < 1.This interpretation,
briefly mentioned in the introduction is reiterated below: }\large }

\paragraph{\textmd{\large ~~~~~~~~Today we have come to regard a QFT describing
physics as an effective field theory arising from a substructure that may become
visible at a much higher energy scale.~ Let us say that this energy scale is
\protect\( \Lambda \protect \) beyond which physics should be described by
a substructure given in terms of a different QFT (valid possibly upto a yet
larger scale).~ We expect that the effective field theory arising out this
high energy QFT should contain the scale \protect\( \Lambda \protect \) in
some manner.~ Here we make a concrete suggestion that this effective theory
is in fact the non-local type theory NLQFT.~ }\large }

\paragraph{\textmd{\large Thus, for example, nonlocalized version of the Standard Model
with a scale \protect\( \Lambda \protect \) (and not the local version) is
being understood as the effective field theory which should break down beyond
scale \protect\( \Lambda \protect \). We wish to propose that the knowledge
of the scale \protect\( \Lambda \protect \) is in fact contained in the NL
formulation of the model and that it is retrieved via the relation (3.1) an
analogue of 0 < Z < 1 (where possible) implemented in perturbation theory. }\large }

\paragraph{\textmd{\large To be concrete, we shall prove the relation :}\large }

\subparagraph{\textmd{\large X( \protect\( \Lambda \protect \)\protect\( ^{2}\protect \))=
Z +~~\protect\( \int _{_{_{_{m^{2}_{1}}}}}^{\infty }\protect \)\protect\( \rho (\sigma \protect \)\protect\( ^{2},\Lambda ^{2},m^{2},\lambda )d\sigma \protect \)\protect\( ^{2}\protect \)~~~~~~~~~~~~~~~~~~~~~~~~~~~~(3.1)}\large }

\subparagraph{\textmd{\large where X( \protect\( \Lambda \protect \)\protect\( ^{2}\protect \))
--> 1~ as~ \protect\( \Lambda \protect \)\protect\( ^{2}\protect \)-->~\protect\( \infty \protect \).}\large }

{\large We shall give a brief derivation for this relation and demonstrate that
, in fact, the derivation can be based} \emph{\large on general principles only}
{\large and without the need for discussion on how the nonlocal theory has to
be quantized.The derivation proceeds more or less along the lines of the corresponding
derivation for the local QFT's; as for example done in {[}6{]} .We shall extensively
draw upon it and point out the essential modifications nonlocality makes. As
is usual for a LSZ derivation, we shall assume Lorentz covariance, existence
of 4-momentum operators {[}translational invariance{]}, positivity of the spectrum
for P\( _{0} \), and completeness of the scattering states which are eigenstates
of the momentum 4-vector: all this we expect in spite of nonlocality. Following
{[}6{]} we consider <0| \( \phi  \)(x)\( \phi  \)(y) |0> and expand in the
basis of the complete set of the scattering states that are eigenstates of P\( _{\mu } \)
with eigenvalue p\( _{_{n_{\mu }}} \).Then use of translational invariance
allows us to write <0|\( \phi  \)(x)|n> = exp(-ip\( _{_{n}} \).x) <0|\( \phi  \)(0)|n>,
we can write {[}6{]}}{\large \par}

{\large <0 |\( \phi  \)(x)\( \phi  \)(y) |0> = \( \frac{1}{(2\pi )^{3}} \)\( \int  \)
d\( ^{4}q \) \( \rho  \)(q) exp \{-iq.(x-y)\} \( \qquad \qquad \qquad  \)(3.2)}{\large \par}

{\large with }{\large \par}

{\large \( \rho  \)(q) \( \equiv  \)(2\( \pi  \))\( ^{3} \)\( \sum _{_{n}} \)
\( \delta  \)\( ^{4} \)(p\( _{_{n}} \)-q) |<0|\( \phi  \)(x)|n>|\( ^{2} \)
\( \qquad \qquad \qquad  \)(3.3)}{\large \par}

{\large The quantity \( \rho  \)(q) is now modified as |<0|\( \phi  \)(x)|n>|
will be modified by nonlocalization; and in fact it will be damped out for large
p\( ^{2}_{_{n}} \) . As in local theory,Lorentz covariance and positivity of
spectrum of P\( _{0} \) implies }{\large \par}

{\large \( \rho  \)(q) =\( \rho  \)(q\( ^{2} \),\( \Lambda  \)\( ^{2} \))
\( \theta  \)(q\( _{0}) \).\( \qquad \qquad \qquad  \)\( \qquad \qquad \qquad  \)(3.4)}{\large \par}

{\large Then (3 .2)--(3.4) allow us to write }{\large \par}

{\large \( \Delta  \)'(x-y) = -i<0|{[} \( \phi  \)(x),\( \phi  \)(y){]} |0>
=\( \frac{-i}{(2\pi )^{3}} \)\( \int  \) d\( ^{4}q \) \( \rho  \)(q\( ^{2} \),\( \Lambda  \)\( ^{2} \))
{[}exp \{-iq.(x-y)\}- exp \{iq.(x-y)\}{]} }{\large \par}

{\large =\( \int  \)d\( \sigma  \)\( ^{2} \) \( \rho  \)(\( \sigma  \)\( ^{2} \),\( \Lambda  \)\( ^{2} \))\( \int  \)d\( ^{4}q \)\( \frac{-i}{(2\pi )^{3}} \)\( \delta  \)(\( \sigma  \)\( ^{2} \)-q\( ^{2} \)){[}exp
\{-iq.(x-y)\}- exp \{iq.(x-y)\}{]} \( \qquad \qquad \qquad  \)(3.5a)}{\large \par}

{\large \( \equiv  \)\( \int  \)d\( \sigma  \)\( ^{2} \) \( \rho  \)(\( \sigma  \)\( ^{2} \),\( \Lambda  \)\( ^{2} \))\( \Delta  \)(x-y,\( \sigma  \)\( ^{2} \))
\( \qquad \qquad \qquad  \)(3.5b)}{\large \par}

{\large where \( \Delta  \)(x-y,\( \sigma  \)\( ^{2} \)) is the corresponding
quantity associated with the} \emph{\large local free field} {\large theory
of mass \( \sigma  \).We now differentiate (3.5b) with respect to x\( _{0} \)
and use the} \emph{\large identity}{\large \par}

{\large \( \frac{\partial }{\partial x_{0}} \)\( \Delta  \)(x-y)\( \Vert  \)\( _{_{x_{0}=y_{0}}} \)
= - \( \delta  \)\( ^{3} \)(\( \mathbf{x}-\mathbf{y}) \)\( \qquad \qquad \qquad  \)(3.6)}{\large \par}

{\large to obtain}{\large \par}

{\large \( \frac{\partial }{\partial x_{0}} \)\( \Delta  \)'(x-y)\( \Vert  \)\( _{_{x_{0}=y_{0}}} \)
= \( \int  \)\( ^{^{\infty }}_{_{_{0}}} \) d\( \sigma  \)\( ^{2} \) \( \rho  \)(\( \sigma  \)\( ^{2} \),\( \Lambda  \)\( ^{2} \))
\( \delta  \)\( ^{3} \)(\( \mathbf{x}-\mathbf{y}) \)}{\large \par}

{\large \( \; \; \; \qquad \qquad \qquad \qquad \qquad \qquad \qquad  \)\( \equiv  \)X{[}\( \Lambda  \)\( ^{2} \),m\( ^{2} \)
{]}\( \delta  \)\( ^{3} \)(\( \mathbf{x}-\mathbf{y}) \) \( \qquad \qquad \qquad  \)(3.7)}{\large \par}

{\large Here, for a finite \( \Lambda  \)\( ^{2} \), X{[}\( \Lambda  \)\( ^{2} \),m\( ^{2} \)
{]} is a finite quantity. Moreover,we know that in the local limit, X{[}\( \Lambda  \)\( ^{2} \),m\( ^{2} \)
{]}\( \rightarrow  \) 1}\footnote{%
This information is all that we need (and have to demand) about the nonlocal
theory: we can do away with details of how the theory is quantized for the present
purpose.
}{\large .Further , Lorentz covariance alone dictates that, for a single particle
state |p>, <0|\( \phi  \)(0)|p> must be of the form {[}recall that with the
present normalization,\( \sqrt{2\omega _{p}} \) |p> is a covariant state, and
the only Lorentz invariant possible is p\( ^{2} \)= m\( ^{2} \) = constant{]}}{\large \par}

{\large <0|\( \phi  \)(0)|p> = \( \frac{1}{\sqrt{(2\pi )^{3}2\omega _{p}}} \)x
(a constant) \( \equiv  \)\( \frac{\sqrt{Z'}}{\sqrt{(2\pi )^{3}2\omega _{p}}} \)
\( \qquad \qquad \qquad  \)(3.8)}{\large \par}

{\large On account of the assumed spectrum of scattering states, \( \rho  \)(\( \sigma  \)\( ^{2} \),\( \Lambda  \)\( ^{2} \))=0
, 0<\( \sigma  \)\( ^{2} \) < m\( ^{2} \). Further, in }{\large \par}

{\large X{[}\( \Lambda  \)\( ^{2} \),m\( ^{2} \) {]}= \( \int  \)\( ^{^{\infty }}_{_{_{0}}} \)
d\( \sigma  \)\( ^{2} \) \( \rho  \)(\( \sigma  \)\( ^{2} \),\( \Lambda  \)\( ^{2} \))
\( \qquad \qquad \qquad  \)(3.9)}{\large \par}

{\large We can evaluate the single particle contribution using (3 .8) as is
done in the local case to obtain}{\large \par}

{\large \( \rho  \)(\( \sigma  \)\( ^{2} \),\( \Lambda  \)\( ^{2} \)) =
Z'\( \delta  \)(\( \sigma  \)\( ^{2} \)-m\( ^{2} \)) + contribution coming
from \( \sigma  \)\( ^{2} \) > m\( ^{2} \). \( \qquad \qquad \qquad  \)(3.10)}{\large \par}

{\large This leads us to}\footnote{%
An analogous exercise carried out for <0| T\{\( \phi  \)(x)\( \phi  \)(y)\}|0>
as in ref.6 allows one to identify Z' with the on-shell renormalization constant.
}{\large \par}

{\large X{[}\( \Lambda  \)\( ^{2} \),m\( ^{2} \) {]}=Z + \( \int  \)\( _{_{_{m^{2}}}} \)\( ^{^{\infty }} \)
d\( \sigma  \)\( ^{2} \) \( \rho  \)(\( \sigma  \)\( ^{2} \),\( \Lambda  \)\( ^{2} \))
\( \qquad \qquad \qquad  \)(3.11)}{\large \par}

{\large with the constraint X{[}\( \Lambda  \)\( ^{2} \),m\( ^{2} \) {]}\( \rightarrow  \)
1.}{\large \par}

\subparagraph{\textmd{\large We now propose a purely theoretical criterion for determining
the maximum allowed value for \protect\( \Lambda \protect \). We shall soon
argue below as to how it is highly plausible. We require that for a given QFT
with given parameters,(an estimate of )the maximum value of \protect\( \Lambda \protect \)
at which the theory should break down as an effective theory is precisely the
one at which Z in (3.11) becomes zero.~ Unlike the usual assumptions about
the QFT, we do not assume that the theory holds beyond scale \protect\( \Lambda \protect \)
and we do not allow for the possibility that Z can be negative.~ }\large }

We {\large shall now present a heuristic argument to show why this picture is
highly plausible.We do it with the help of a simple model.Let us imagine that
at low energies {[}E <\textcompwordmark{}<E\( _{0} \){]} ,physics is described
by the scalar field theory presented here;and that the scalar is a bound state
of a fermion-antifermion pair {[}\( \psi  \),\( \overline{\psi } \){]}.At
high energies {[}E>\textcompwordmark{}>E\( _{0} \){]}, we expect the theory
to be replaced by a QFT of \( \psi  \)'s. We consider the composite operator
\( \overline{\psi } \)(x)\( \psi  \)(x) as the interpolating field operator
for \( \phi  \).We now consider \( \overline{\psi } \)(0)\( \psi  \)(0) |0>
for both E <\textcompwordmark{}<E\( _{0} \) and E>\textcompwordmark{}>E\( _{0} \).}{\large \par}

{\large \( \overline{\psi } \)(0)\( \psi  \)(0) |0> = \( \sum  \)\( _{_{n}} \)
a\( _{n} \) |n,p\( _{n}> \); for E <\textcompwordmark{}<E\( _{0} \)~~~~~~~~~~~~~~~~~~
(3.12a)}{\large \par}

{\large \( \overline{\psi } \)(0)\( \psi  \)(0) |0> = \( \sum  \)\( _{_{n}} \)
a\( _{n} \)' |n,p\( _{n}> \)' ;for E>\textcompwordmark{}>E\( _{0} \)~~~~~~~~~~~~~~~~~~
(3.12b) }{\large \par}

{\large In (3.12a), the states |n,p\( _{n}> \) refer to the eigenstates of
the momentum operator that include the single scalar particle states and the
states corresponding to the scattering of scalars , whereas in (3.12b), |n,p\( _{n}> \)'
includes these} \emph{\large (}{\large directly or indirectly)} \emph{\large and
all possible states containing fermions and antifermions and these include scattering
states of these.} {\large Clearly, (3.12b) contains too many more states {[}
more channels open up, this includes effects of much more phase space{]} than
(3.12a). So the} \emph{\large relative} {\large probability for finding a single
scalar particle state in such an expansion, which is related to Z in the scalar
theory, has a sudden fall around E=E\( _{0} \). We should thus expect that
Z in the scalar theory should fall}\footnote{%
We are using the full fermion theory to argue this out; however, we may expect
a similar conclusion to hold qualitatively in the effective theory.
} {\large to a small value around E=E\( _{0} \) where the effective theory should
be replaced by a more fundamental fermion theory}\footnote{%
We assume that this fermion theory in turn is a NLQFT with a higher parameter
\( \Lambda  \)' >\textcompwordmark{}>\( \Lambda  \);so that such probabilities
continue to be well-defined in perturbation theory.
}{\large .}{\large \par}

\paragraph{\textmd{\large We, thus, expect a condition of the \char`\"{}form\char`\"{}:}\large }

\paragraph{\textmd{\large \protect\( \frac{\lambda }{^{^{_{16\pi ^{2}}}}}ln\protect \)
\protect\( \frac{\Lambda ^{2}}{_{_{m^{2}}}}\protect \)~< 1~~~~~~~~~~~~~~~~~~~~~~~~~~~~~~~~~~~~~~~~~~~~~~~~~~~~~~~~~~~~~~~~~~~
(3.13)}\large }

\paragraph{\textmd{\large This yields, (without~ worrying about exact coefficients in
(3.13) )}\large }

\paragraph{\textmd{\large \protect\( \Lambda \protect \)\protect\( _{max}\protect \)~~
= m exp \{ 8 \protect\( \pi ^{2}\protect \)/ \protect\( \lambda \protect \)\}}\large }

\paragraph{\textmd{\large For m = 1 GeV, and~~ \protect\( \lambda \protect \)/16 \protect\( \pi ^{2}\protect \)
= 0.05 {[}0.01{]}~ we obtain:}\large }

\paragraph{\textmd{\large \protect\( \Lambda \protect \)\protect\( _{max}\protect \)
= 22 TeV {[} 10\protect\( ^{18}\protect \) TeV{]}.}\large }

\paragraph{\textmd{\large Of course, the actual numbers are sensitive to the coefficient
in (3.13) and to the value of \protect\( \lambda \protect \) in a given theory;however
we may expect abound that is testable in near future. }\large }

\paragraph{\textmd{\large ~}\large }

\end{document}